\documentclass[a4paper,11pt]{article}
\usepackage{pos}

\title{The Direct Search Experiment for Light Dark Matter
(DELight): Overview and Perspectives}

\author*[a]{Melih Solmaz}
\onbehalf{on behalf of the DELight collaboration}

\affiliation[a]{Kirchhoff-Institute for Physics, Heidelberg University, 69120 Heidelberg, Germany}

\emailAdd{melih.solmaz@kip.uni-heidelberg.de}

\abstract{Driven by the null results in the searches for dark matter,
the field of direct dark matter detection is constantly evolving
to push new frontiers. Ultimately, a vast parameter space for dark matter
masses below a few GeV is yet to be explored. That said, low mass dark matter
candidates necessitate novel detector designs with lower thresholds and
alternative target materials.
The Direct search Experiment for Light dark matter (DELight) will deploy a
target of superfluid $^4$He instrumented with large area microcalorimeters
(LAMCALs) based on magnetic microcalorimeter (MMC) technology in a setup
optimized for low mass dark matter searches. This paper presents
an overview of this novel upcoming experiment, including
detection technology, sensitivity goals and R\&D studies.}

\FullConference{19th International Conference on Topics in Astroparticle and Underground Physics (TAUP2025)\\
24–30 Aug 2025\\
Xichang, China\\}

\begin{document}
\maketitle

\section{Motivation}
\label{sec:motiv}

No compelling evidence of dark matter (DM)-induced signal detection
has been observed to date but significant progress has been made in probing the parameter space for thermally produced DM. The following points encapsulate the current landscape of DM direct detection
searches~\cite{RPP2024}: (i) The best limits were set by
liquid xenon experiments~\cite{LZ:2022lsv, XENON:2023cxc, PandaX-4T:2021bab}, which target DM masses between 10\,GeV and 1\,TeV. (ii) Several experiments improved the limits at GeV and sub-GeV masses thanks to
their low threshold~\cite{CRESST:2019jnq, DarkSide-50:2022qzh, XENON:2020gfr}.
(iii) Despite these achievements, a significant portion of the
sub-GeV DM parameter space has not yet been explored, motivating experimental
efforts to focus on the search for light DM (LDM) particles~\cite{Knapen:2017xzo}.

Direct detection experiments commonly hunt for nuclear recoil (NR)
signals caused by the elastic scattering of DM particles. The NR
spectrum varies with the DM and target nucleus masses.
For a 1\,GeV LDM scenario, the NR spectra on the
currently used targets, such as xenon and silicon, remain confined to
recoil energies $\leq$0.5\,keV.
However, considering helium, the upper bound of the NR spectrum is
slightly above 2\,keV.
Specifically, a low-mass target such as helium provides a more efficient
momentum transfer, thereby enhancing the sensitivity to LDM.

The advantage of $^4$He as a potential target is not limited to the
recoil energy enhancement. It is conveniently scalable, it has no intrinsic long-lived isotopes, 
and it is self-purified
at superfluid temperatures as the contaminants are frozen out. Furthermore,
superfluid helium produces multiple distinct signals following an energy
deposit~\cite{DELight:2024bgv}. The energy is distributed among three
channels, namely heat, excitation, and ionization. Heat generates
phonons and rotons, jointly referred to as quasiparticles.
Excitation mostly creates infrared (IR) photons, singlet and
triplet helium excimers. Singlet states decay almost instantly,
producing ultraviolet (UV) photons. The triplet state has a long
lifetime of about 13\,s, producing delayed UV scintillation. In the case
of ionization under zero electric field, singlets
(prompt UVs) and triplets (delayed UVs) are
generated upon recombination.

Based on these attractive features,
superfluid $^4$He was previously proposed as a target material
for LDM searches~\cite{Lanou:1988iq, Adams:1996ge}.
The Direct search Experiment for Light Dark Matter, DELight,
is a proposed superfluid $^4$He-based particle detector to search for
elusive NR signals induced by sub-GeV DM particles~\cite{vonKrosigk:2022vnf}.
The DELight Collaboration is currently laying the groundwork
for this novel LDM direct detection technology and is
composed of three German institutions:
Heidelberg University, Karlsruhe Institute of Technology, and University of Freiburg.

\section{DELight Detector}
\label{sec:detec}

\begin{figure}[htbp]
\centering
\includegraphics[width=\textwidth]{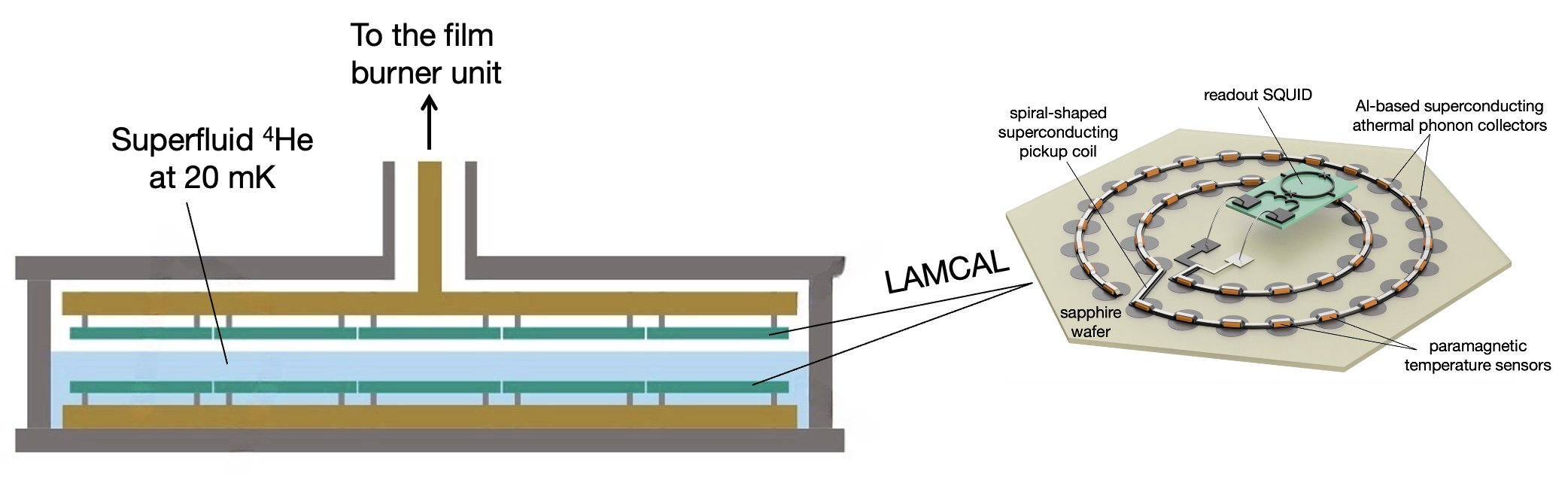}
\caption{Sketch of the DELight Phase-I detector. Left: Pancake geometry of the helium cell with top and bottom LAMCAL array. Right: Schematic of a large area microcalorimeter (LAMCAL).}
\label{fig:Detector}
\end{figure}

Figure~\ref{fig:Detector} illustrates the DELight cell for Phase-I of the experiment,
which has a pancake-type geometry. The cell is equipped with two
wafer calorimeter arrays, one submerged in liquid and the other in vacuum. As can be
inferred from Section~\ref{sec:motiv}, the signal generated
in the superfluid $^4$He has essentially two components: scintillation
and quasiparticles. The scintillation signal consists of prompt
IR photons and prompt and delayed UV photons. The liquid
medium is entirely transparent to these photons, which can thus reach both the top and bottom sensor array.
No scintillation occurs below 19.8\,eV (i.e., the first excited state energy of He).

Regarding quasiparticles, they propagate ballistically inside helium
at a speed of 150-200\,m/s. Thus, they produce slow signals as opposed to
the scintillation. The phenomenon of ``quantum evaporation'' plays a central
role in the detection of quasiparticles~\cite{DELight:2024bgv}.
When a quasiparticle reaches the vacuum interface with an energy
greater than the helium binding energy (0.62\,meV) to the liquid
surface, it can evaporate a single $^4$He atom into the
vacuum. The evaporated atom travels to one of the wafer calorimeters
in the top array and is adsorbed onto the wafer
surface. The adsorption releases $\mathcal{O}(10)$\,meV of
material-dependent energy that is identical to the
binding energy of helium to the wafer material. Consequently,
signal amplification $\geq$10 is accomplished.
It should be emphasized that the pancake
design maximizes the quantum evaporation efficiency as shown by simulations,
enabling access to a nuclear recoil energy threshold notably below 100\,eV.

As an inherent property, superfluid helium creeps up along the inner walls of its container in the form of a thin film. Importantly,
the aforementioned amplification gain is attained, only if
the surface of the top sensor array is kept He-film free.
Failing to do so would decrease the gain. To address this issue,
we will instrument a film burner unit on top of
the cell~\cite{Torii1992, Adams:1999yk}.
The fundamental aspect of this system is to restrict the flow of
the helium film down to the top wafer calorimeters through evaporation of helium atoms,
which ascend along the walls toward the film burner.

The sensitive top and bottom arrays will be a collection of large area microcalorimeters
(LAMCALs). The inset of figure~\ref{fig:Detector} shows what constitutes
each LAMCAL. LAMCALs are assembled on sapphire wafers. When UV photons or quasiparticles deposit energy in the wafer, phonons are generated and subsequently collected by dedicated phonon collectors.
This leads to a temperature increase in the paramagnetic
sensors, magnetic microcalorimeters (MMCs)~\cite{Kempf2018}. MMCs
transduce the temperature rise into a change of magnetization,
which is read out by a superconducting quantum interference device (SQUID).
DELight members made exceptional advancements in this technology.
They demonstrated that MMCs, designed specifically for x-ray detection,
achieve a world-leading full width at half maximum (FWHM) resolution
of 1.25\,eV at 5.9\,keV~\cite{Krantz2024, Toschi:2023spv}.
As a baseline scenario, the DELight LAMCALs are projected to have a
FWHM baseline resolution of 2.5\,eV and an energy threshold of $\sim$5\,eV, equivalent to about 5$\sigma$ baseline resolution.

\section{From Foundation to Expansion: DELight Phase I-II}

In phase-I, DELight will be located at the Vue-des-Alpes (VdA) underground
laboratory~\cite{Gonin2003}, a shallow lab in Switzerland
with 230\,m of rock overburden ($\sim$630\,mwe). It hosts
a low-background facility for gamma spectrometry, GeMSE~\cite{Garcia:2022jdt},
which is operated by collaboration members. Radioassay of various potential DELight materials with GeMSE has started.

Alongside this, a dedicated simulation framework has also been developed
to provide the foundational basis for the DELight experiment.
A preliminary geometry was implemented in Geant4, including the
dilution refrigerator and the film burner. As part of simulation
efforts, the Collaboration originated a Monte Carlo formalism that
describes signal formation and partitioning in superfluid $^4$He~\cite{DELight:2024bgv}.
The work demonstrates that the target has a great potential to differentiate NR events from electronic recoil (ER) events.
Moreover, specific quasiparticle characteristics---such as their complex
dispersion relations, quantum evaporation, and interactions with solid
surfaces~\cite{Hertel:2018aal}---were incorporated into the Geant4 simulations. The simulation framework enables the generation of
NR/ER event traces across a range of energies
for a DELight setup with LAMCAL readout.

\begin{figure}[htbp]
\centering
\includegraphics[width=0.46\textwidth]{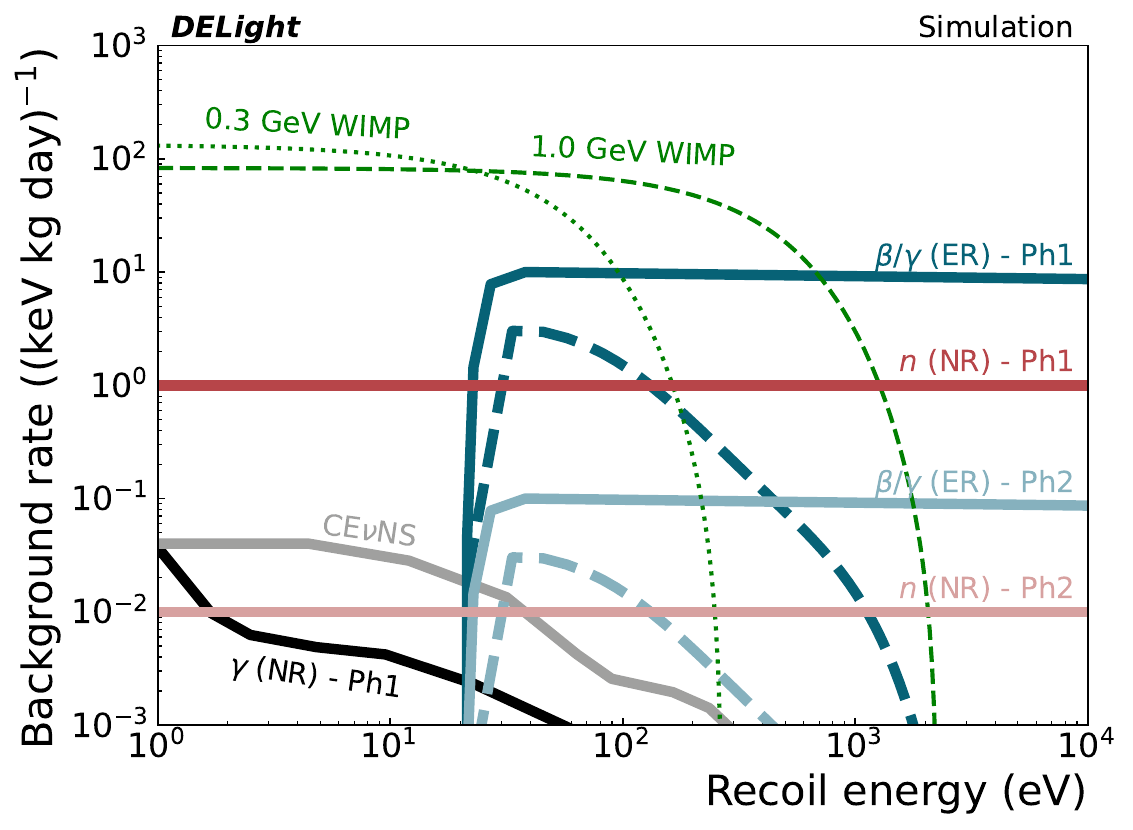}
\includegraphics[width=0.52\textwidth]{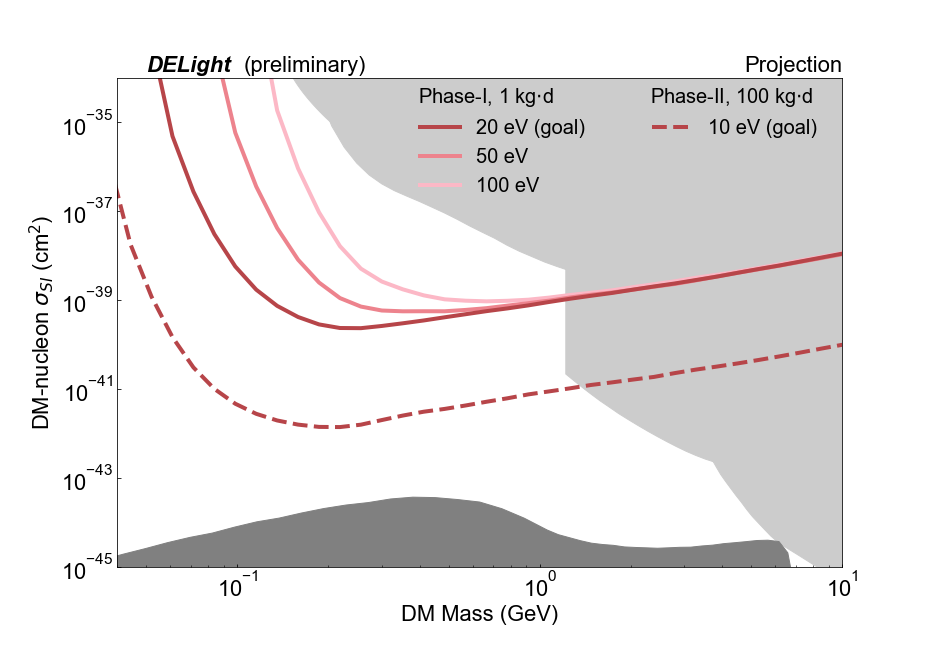}
\caption{(Left) DELight background goals for phase-I/II (solid lines). The dashed
dark(light) blue line is the phase-I(II) ER goal background
after a rudimentary ER/NR discrimination cut, drawn for instructive
purposes.
(Right) Sensitivity projections at 90\% CL for DELight phase-I/II at the goal background-levels without ER/NR discrimination.}
\label{fig:DELightGoals}
\end{figure}

To determine the experimental sensitivity, we established a preliminary yet comprehensive background model, taking radiogenics owing to the detector components and
the laboratory, solar neutrinos, photonuclear processes and muon-induced
events into account. The simulations were conducted without
incorporating any external passive shield against the lab background.
As illustrated in figure~\ref{fig:DELightGoals} (left),
the ER(NR) background goal for the first phase of DELight
is 10(1)\,events/(keV$\cdot$kg$\cdot$day). Simulations suggest that
this goal can be reached through moderate shielding, providing a factor
of 10$^3$ reduction of laboratory background. Figure~\ref{fig:DELightGoals} (left)
also indicates that the NR background rates from solar
neutrinos (gray) and photonuclear events (black) can be
afforded for phase-I. The phase-II background goals are two orders
of magnitude lower in relation to phase-I.

The energy threshold goal of DELight is 20\,eV for phase-I.
The preliminary simulations show that if the baseline LAMCAL energy
resolution, stated in section~\ref{sec:detec}, is achieved,
this threshold goal is a conservative assumption.
During phase-I, DELight is expected to run at this threshold
with 1\,L of superfluid $^4$He target to reach 1\,kg$\cdot$day
exposure. The phase-I sensitivity goal is shown
in figure~\ref{fig:DELightGoals} (right). Additionally,
figure~\ref{fig:DELightGoals} (right) implies that even more conservative
thresholds of 50\,eV and 100\,eV will produce competitive limits.
For phase-II, a target volume of 10\,L is planned. The objective is
to produce a front-running limit, as illustrated in
figure~\ref{fig:DELightGoals} (right), at an energy threshold of
10\,eV with an exposure of 100\,kg$\cdot$day.

\section{DELight Demonstrator}

\begin{figure}[htbp]
\centering
\includegraphics[width=0.24\textwidth]{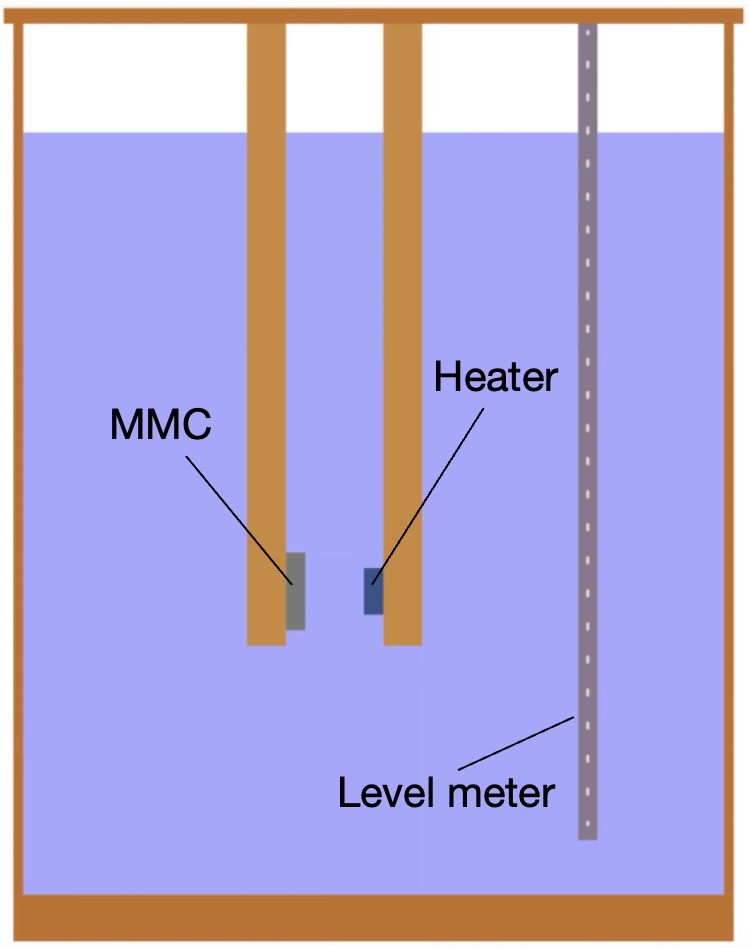}\hspace{35pt}
\includegraphics[width=0.54\textwidth]{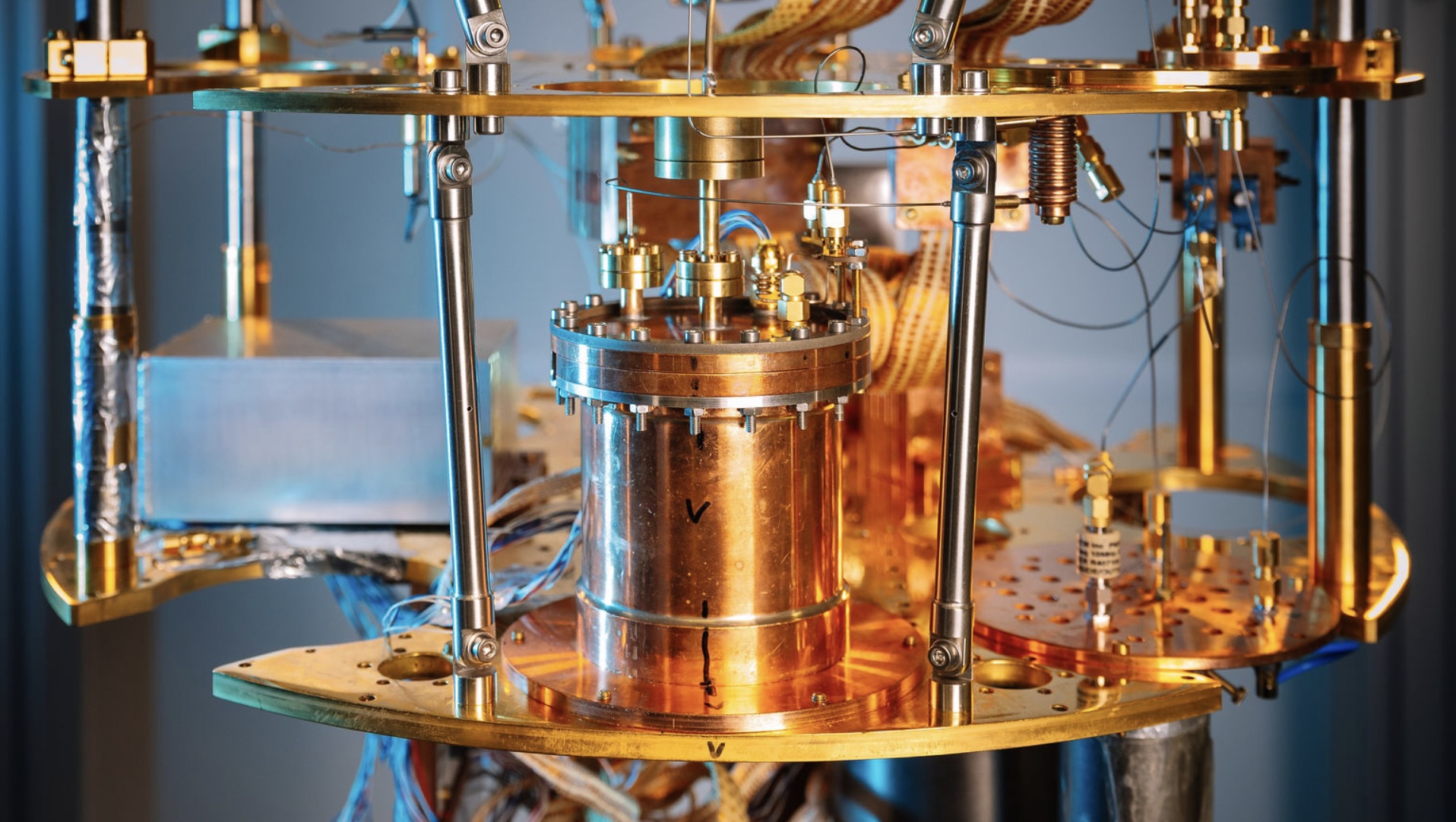}
\caption{(Left) Interior of the DELight Demonstrator cell filled with  $\sim$300\,ml helium (purple). (Right) Picture of the DELight Demonstrator attached to the dilution refrigerator. Image credit:
Florian Freundt.}
\label{fig:Demonstrator}
\end{figure}

The Collaboration assembled an R\&D helium cell to test and demonstrate the functioning of key parts of DELight, which is called ``DELight Demonstrator''. It is
a cell holding $\sim$300\,ml equipped with an MMC and a heater, as
illustrated in figure~\ref{fig:Demonstrator} (left).
The Demonstrator enables among other things testing the level meter and the study of the sensor behavior in
superfluid $^4$He along with the development of pulse recognition and
event reconstruction methods. It also serves as a platform to investigate
the possibility of measuring quasiparticle signals in liquid
through monitoring the sensor response to heater pulses.
The transmission of quasiparticles into the submerged MMC is expected to be strongly suppresed by the extremely large Kapitza resistance~\cite{Hertel:2018aal}.
Until now, the Collaboration constructed and commissioned the leak-tight cell,
as shown in figure~\ref{fig:Demonstrator} (right) and operates
it with superfluid $^4$He at $\sim$15\,mK. The analysis of
the first measurements with the immersed MMC is presently underway.

\section{Conclusion}

Motivated by the null results in DM searches, the domain of the DM direct
detection is pushed towards sub-GeV mass scales.
The Direct search Experiment for Light Dark Matter, DELight, is a
proposed experiment to search for elusive NR signals induced by
LDM particles. DELight employs a superfluid $^4$He cell
equipped with LAMCAL arrays to detect UV scintillation and
quasiparticles as a result of particle interactions.
In its phase-I (1\,L of superfluid $^4$He),
DELight aims to reach an energy threshold of
20\,eV and produce a world-leading LDM search limit with an exposure of
1\,kg$\cdot$day. Initial studies show that DELight will still yield
competitive limits even for moderate threshold scenarios of 50\,eV and 100\,eV.
Furthermore, an R\&D cell, the DELight Demonstrator, was assembled
and successfully commissioned to establish the proof-of-principle
for several key aspects of the experiment.

\acknowledgments{This work was supported by the Heidelberg Karlsruhe Strategic Partnership (HEiKA STAR).
We also acknowledge
the support through the Alexander von Humboldt Foundation.}

\bibliographystyle{JHEP}
\bibliography{references}

@article{RPP2024,
  title = {Review of Particle Physics},
  author = "Navas, S. and others",
  collaboration = {Particle Data Group Collaboration},
  journal = {Phys. Rev. D},
  volume = {110},
  issue = {3},
  pages = {030001},
  numpages = {5},
  year = {2024},
  month = {Aug},
  publisher = {American Physical Society},
  doi = {10.1103/PhysRevD.110.030001},
  url = {https://link.aps.org/doi/10.1103/PhysRevD.110.030001}
}

@article{XENON:2023cxc,
    author = "Aprile, E. and others",
    collaboration = "XENON",
    title = "{First Dark Matter Search with Nuclear Recoils from the XENONnT Experiment}",
    eprint = "2303.14729",
    archivePrefix = "arXiv",
    primaryClass = "hep-ex",
    doi = "10.1103/PhysRevLett.131.041003",
    journal = "Phys. Rev. Lett.",
    volume = "131",
    number = "4",
    pages = "041003",
    year = "2023"
}

@article{LZ:2022lsv,
    author = "Aalbers, J. and others",
    collaboration = "LZ",
    title = "{First Dark Matter Search Results from the LUX-ZEPLIN (LZ) Experiment}",
    eprint = "2207.03764",
    archivePrefix = "arXiv",
    primaryClass = "hep-ex",
    doi = "10.1103/PhysRevLett.131.041002",
    journal = "Phys. Rev. Lett.",
    volume = "131",
    number = "4",
    pages = "041002",
    year = "2023"
}

@article{PandaX-4T:2021bab,
    author = "Meng, Yue and others",
    collaboration = "PandaX-4T",
    title = "{Dark Matter Search Results from the PandaX-4T Commissioning Run}",
    eprint = "2107.13438",
    archivePrefix = "arXiv",
    primaryClass = "hep-ex",
    doi = "10.1103/PhysRevLett.127.261802",
    journal = "Phys. Rev. Lett.",
    volume = "127",
    number = "26",
    pages = "261802",
    year = "2021"
}

@article{CRESST:2019jnq,
    author = "Abdelhameed, A. H. and others",
    collaboration = "CRESST",
    title = "{First results from the CRESST-III low-mass dark matter program}",
    eprint = "1904.00498",
    archivePrefix = "arXiv",
    primaryClass = "astro-ph.CO",
    doi = "10.1103/PhysRevD.100.102002",
    journal = "Phys. Rev. D",
    volume = "100",
    number = "10",
    pages = "102002",
    year = "2019"
}

@article{DarkSide-50:2022qzh,
    author = "Agnes, P. and others",
    collaboration = "DarkSide-50",
    title = "{Search for low-mass dark matter WIMPs with 12~ton-day exposure of DarkSide-50}",
    eprint = "2207.11966",
    archivePrefix = "arXiv",
    primaryClass = "hep-ex",
    reportNumber = "FERMILAB-PUB-22-589-ND-PPD-SCD",
    doi = "10.1103/PhysRevD.107.063001",
    journal = "Phys. Rev. D",
    volume = "107",
    number = "6",
    pages = "063001",
    year = "2023"
}

@article{XENON:2020gfr,
    author = "Aprile, E. and others",
    collaboration = "XENON",
    title = "{Search for Coherent Elastic Scattering of Solar $^8$B Neutrinos in the XENON1T Dark Matter Experiment}",
    eprint = "2012.02846",
    archivePrefix = "arXiv",
    primaryClass = "hep-ex",
    doi = "10.1103/PhysRevLett.126.091301",
    journal = "Phys. Rev. Lett.",
    volume = "126",
    pages = "091301",
    year = "2021"
}

@article{Knapen:2017xzo,
    author = "Knapen, Simon and Lin, Tongyan and Zurek, Kathryn M.",
    title = "{Light Dark Matter: Models and Constraints}",
    eprint = "1709.07882",
    archivePrefix = "arXiv",
    primaryClass = "hep-ph",
    doi = "10.1103/PhysRevD.96.115021",
    journal = "Phys. Rev. D",
    volume = "96",
    number = "11",
    pages = "115021",
    year = "2017"
}

@article{DELight:2024bgv,
    author = "Toschi, Francesco and others",
    collaboration = "DELight",
    title = "{Signal partitioning in superfluid He4: A Monte~Carlo approach}",
    eprint = "2410.13684",
    archivePrefix = "arXiv",
    primaryClass = "hep-ex",
    doi = "10.1103/PhysRevD.111.032013",
    journal = "Phys. Rev. D",
    volume = "111",
    number = "3",
    pages = "032013",
    year = "2025"
}

@inproceedings{Lanou:1988iq,
    author = "Lanou, R. E. and Maris, H. J. and Seidel, G. M.",
    title = "{SUPERFLUID HELIUM AS A DARK MATTER DETECTOR}",
    booktitle = "{23rd Rencontres de Moriond: Astronomy: Detection of Dark Matter}",
    pages = "75--83",
    year = "1988"
}

@inproceedings{Adams:1996ge,
    author = "Adams, J. S. and Bandler, S. R. and Brouer, S. M. and Enss, C. and Lanou, R. E. and Maris, H. J. and More, T. and Seidel, G. M.",
    title = "{'HERON' as a dark matter detector?}",
    booktitle = "{31st Rencontres de Moriond: Dark Matter and Cosmology, Quantum Measurements and Experimental Gravitation}",
    pages = "131--136",
    year = "1996"
}

@article{vonKrosigk:2022vnf,
    author = "von Krosigk, Belina and others",
    title = "{DELight: A Direct search Experiment for Light dark matter with superfluid helium}",
    eprint = "2209.10950",
    archivePrefix = "arXiv",
    primaryClass = "hep-ex",
    doi = "10.21468/SciPostPhysProc.12.016",
    journal = "SciPost Phys. Proc.",
    volume = "12",
    pages = "016",
    year = "2023"
}

@article{Torii1992,
    author = {Torii, R. and Bandler, S. R. and More, T. and Porter, F. S. and Lanou, R. E. and Maris, H. J. and Seidel, G. M.},
    title = {Removal of superfluid helium films from surfaces below 0.1 K},
    journal = {Review of Scientific Instruments},
    volume = {63},
    number = {1},
    pages = {230-234},
    year = {1992},
    month = {01},
    issn = {0034-6748},
    doi = {10.1063/1.1142964},
    url = {https://doi.org/10.1063/1.1142964},
}

@article{Adams:1999yk,
    author = "Adams, J. S. and Fleischmann, A. and Huang, Y. H. and Kim, Y. H. and Lanou, R. E. and Maris, H. J. and Seidel, G. M.",
    editor = "Diwan, M. V. and Jung, C. K.",
    title = "{Progress on HERON: A real-time detector for p - p solar neutrinos}",
    doi = "10.1063/1.1361731",
    journal = "AIP Conf. Proc.",
    volume = "533",
    number = "1",
    pages = "112--117",
    year = "2000"
}

@ARTICLE{Kempf2018,
       author = {{Kempf}, S. and {Fleischmann}, A. and {Gastaldo}, L. and {Enss}, C.},
        title = "{Physics and Applications of Metallic Magnetic Calorimeters}",
      journal = {Journal of Low Temperature Physics},
         year = 2018,
        month = nov,
       volume = {193},
       number = {3-4},
        pages = {365-379},
          doi = {10.1007/s10909-018-1891-6},
       adsurl = {https://ui.adsabs.harvard.edu/abs/2018JLTP..193..365K}
}

@article{Krantz2024,
    author = {Krantz, Matthäus and Toschi, Francesco and Maier, Benedikt and Heine, Greta and Enss, Christian and Kempf, Sebastian},
    title = {Magnetic microcalorimeter with paramagnetic temperature sensors and integrated dc-SQUID readout for high-resolution x-ray emission spectroscopy},
    journal = {Applied Physics Letters},
    volume = {124},
    number = {3},
    pages = {032601},
    year = {2024},
    month = {01},
    issn = {0003-6951},
    doi = {10.1063/5.0180903},
    url = {https://doi.org/10.1063/5.0180903}
}

@article{Toschi:2023spv,
    author = "Toschi, Francesco and Maier, Benedikt and Heine, Greta and Ferber, Torben and Kempf, Sebastian and Klute, Markus and von Krosigk, Belina",
    title = "{Optimum filter-based analysis for the characterization of a high-resolution magnetic microcalorimeter}",
    eprint = "2310.08512",
    archivePrefix = "arXiv",
    primaryClass = "hep-ex",
    doi = "10.1103/PhysRevD.109.043035",
    journal = "Phys. Rev. D",
    volume = "109",
    number = "4",
    pages = "043035",
    year = "2024"
}

@article{Gonin2003,
    author = {Gonin, Yvan and Busto, José and Vuilleumier, Jean-Luc},
    title = {The “La Vue-des-Alpes” underground laboratory},
    journal = {Review of Scientific Instruments},
    volume = {74},
    number = {11},
    pages = {4663-4666},
    year = {2003},
    month = {11},
    issn = {0034-6748},
    doi = {10.1063/1.1614854},
    url = {https://doi.org/10.1063/1.1614854}
}

@article{Garcia:2022jdt,
    author = "Garc{\'\i}a, Diego Ram{\'\i}rez and Baur, Daniel and Grigat, Jaron and Hofmann, Beda A. and Lindemann, Sebastian and Masson, Darryl and Schumann, Marc and von Sivers, Moritz and Toschi, Francesco",
    title = "{GeMSE: a low-background facility for gamma-spectrometry at moderate rock overburden}",
    eprint = "2202.06540",
    archivePrefix = "arXiv",
    primaryClass = "physics.ins-det",
    doi = "10.1088/1748-0221/17/04/P04005",
    journal = "JINST",
    volume = "17",
    number = "04",
    pages = "P04005",
    year = "2022"
}

@article{Hertel:2018aal,
    author = "Hertel, S. A. and Biekert, A. and Lin, J. and Velan, V. and McKinsey, D. N.",
    title = "{Direct detection of sub-GeV dark matter using a superfluid $^4$He target}",
    eprint = "1810.06283",
    archivePrefix = "arXiv",
    primaryClass = "physics.ins-det",
    doi = "10.1103/PhysRevD.100.092007",
    journal = "Phys. Rev. D",
    volume = "100",
    number = "9",
    pages = "092007",
    year = "2019"
}

\end{document}